\documentstyle[multicol,aps,prl,epsf]{revtex}

\begin{document}

\draft

\title{Effect of bonding of a CO molecule on the conductance of atomic metal wires}

\author{M. Kiguchi, D. Djukic, and J.M. van Ruitenbeek}

\address{Kamerlingh Onnes Laboratorium, Universiteit Leiden, Postbus 9504, NL - 2300 RA Leiden, The Netherlands}

\date{\today}

\maketitle

\begin{abstract}

We have measured the effect of bonding of a CO molecule on the conductance of Au, Cu, Pt, and Ni atomic contacts 
at 4.2 K. When CO gas is admitted to the metal nano contacts, a conductance feature appears in the conductance 
histogram near 0.5 of the quantum unit of conductance, for all metals. For Au, the intensity of this fractional 
conductance feature can be tuned with the bias voltage, and it disappears at high bias voltage (above $\sim$ 200 mV). The 
bonding of CO to Au appears to be weakest, and associated with monotomic Au wire formation.

\end{abstract}

\medskip

\begin{multicols}{2}
\narrowtext

\section{INTRODUCTION}
\label{sec1}

Fabrication and characterization of nano structures have attracted wide attention because of its potential 
application in atomic scale electronic devices. The development of fabrication methods such as mechanically 
controllable break junctions (MCBJ) and a scanning tunnelling microscopy, has made it possible to fabricate 
even monatomic metal wires \cite{1,2}. The properties of the nano wires have been studied in detail by various 
methods \cite{3}. On the basis of the understanding of the metal nano wire, the influence of molecules binding to 
the metal nano contacts has been discussed in recent years \cite{4,5,6,7,8,9,10,11,12}. Li et al. reported a change in the 
conductance of metal nano wires due to adsorbates. The change was attributed to adsorbate scattering and 
rearrangement in the atomic configuration of the nano wire induced by adsorption \cite{4}. Under specific 
circumstances a single molecule can be made to form a bridge between metal electrodes just after breaking 
the nano contact. The first conductance measurement of a single molecule was performed by Reed et al. who 
investigated 1,4 benzendithiol \cite{5}. Conductance measurements of single molecules have been studied by 
several experimental techniques, including MCBJ and electromigration break junctions \cite{9}, and various 
interesting phenomena have been observed \cite{6,9}. Although these results are promising there are large 
discrepancies between experimental results obtained by different groups, and between experiments and 
theory. One of the reasons for these discrepancies is uncertainty regarding the configuration of the molecule 
and the nature of the molecule-metal interface. The molecules selected for these studies are usually 
composed of several carbo-hydride rings and are anchored to gold metal leads by sulphur groups. In view of 
the difficulties, it seems natural to step back and focus on even simpler system, such as H$_{2}$ and CO.   

For H$_{2}$ on Pt nano contacts detailed studies have been performed by conductance measurement and 
point contact spectroscopy. It was found that the conductance of a single H$_{2}$ molecule anchored between Pt 
leads is slightly below 1 $G_{0}$, where $G_{0}$ = $2e^{2}/h$ is the quantum conductance unit. The configuration of the H$_{2}$ 
molecule in the junction and the number of conductance channels were determined by combined 
experimental and theoretical studies \cite{7,8}. The effect of H$_{2}$ on metal nano contacts was also studied for Au, 
Fe, Co, Ni, and Pd \cite{10,11,12}. Going beyond the simplest molecule, H$_{2}$, we concentrate here on CO for the 
follows reasons. First, CO is one of the simplest molecules for which atomic and electronic structures for 
bonding to metal surfaces are well determined. CO on the metal surfaces can be studied by conventional 
surface analysis methods, including Auger electron spectroscopy (AES) \cite{13} and low energy electron 
diffraction (LEED) \cite{14}, while H$_{2}$ is hard to detect by these methods \cite{15}. We can discuss the results for the 
metal nano contacts based on the understanding obtained for flat metal surfaces. Second, both electron 
donation and back donation from the CO molecule occur at the same time, which are important charge 
transfer mechanisms for the understanding of chemical interactions between a molecule and a metal. In case 
of CO, the 5$\sigma$ orbital (which is the highest occupied molecular orbital: HOMO) and 2$\pi^{*}$ orbital (lowest 
unoccupied molecular orbital: LUMO) stay near the Fermi level of the metal, involving electron donation 
from the 5$\sigma$ orbital to the metal, and back donation from the metal d band to the 2$\pi^{*}$ orbital. Third, CO is a 
poisonous gas. If the conductance of the metal nano contact is changed in a specific way by CO, it may be 
applicable as CO gas sensors.

The effect of CO on the Pt nano contacts is discussed in more detail elsewhere \cite{11,16}. The CO single 
molecule anchored between Pt leads was characterized by conductance measurement and point contact 
spectroscopy. In the present study, we report on the effect of CO on Ni, Cu, Pt, and Au nano contacts. While 
Au and Cu are noble metals, Ni and Pt are transition metals, for which the Fermi level lays inside the metal d 
band and back donation from the metal to the CO 2$\pi^{*}$ orbital should play an important role. Reflecting the 
change in the chemical interaction the adsorption energy of CO on Ni, Pt, Cu, and Au flat surfaces decreases 
in that order \cite{15}. We discuss the relationship between this adsorption character on macroscopic metal 
surfaces and the conductance behaviour of metal nano contacts. For Au nano contacts we present a more 
detailed analysis of the conductance histograms and conductance traces.

\section{EXPERIMENTAL}
\label{sec2}

The measurements have been performed using the mechanically controllable break junction technique 
\cite{3}. A small notch was cut at the middle of Ni, Cu, Pt, and Au wires in order to fix the breaking point. The 
wires used were 0.1 mm in diameter, about 1 cm long. The wire was glued on top of a bending beam and 
mounted in a three-point bending configuration inside a vacuum chamber. Once under vacuum and cooled to 
4.2 K the wire was broken by mechanical bending of the substrate. Clean fracture surfaces are exposed by 
breaking and these remain clean for days in the cryogenic vacuum. The bending can be relaxed to form 
atomic-sized contacts between the wire ends using a piezo element for fine adjustment. CO was admitted via 
a home made capillary equipped with a heater wire running along the capillary that prevents premature 
condensation of the CO gas. After admitting a small amount of CO gas in the sample chamber we waited 
some time for the gas to diffuse to the end of the insert. About 3000 digitized conductance traces were used 
to build each conductance histogram in the present study.

\section{RESULTS}
\label{sec3}

\begin{figure}
\begin{center}
\leavevmode\epsfxsize=65mm \epsfbox{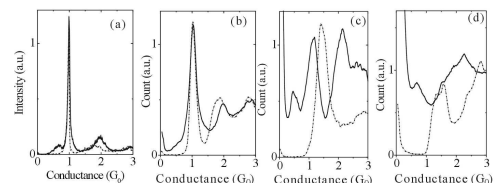}
\caption{
Conductance histograms for (a) Au, (b) Cu, (c) Pt and (d) Ni atomic-size contacts before (dotted-line) 
and after (thick line) admitting CO. The bias voltage is (a) 0.03 V, (b) 0.05 V, (c) 0.15 V and (d) 0.2 V.}
\label{fig1}
\end{center}
\end{figure}

Figure~\ref{fig1} shows the conductance histograms for Ni, Cu, Pt, and Au nano contacts before and after 
admitting CO. The conductance histograms are observed to change by admitting CO, most dramatically for 
Pt and Ni. For Pt, two new peaks appear in the conductance histogram, one near 0.5 $G_{0}$ and one near 1 $G_{0}$, in 
agreement with our previous studies \cite{11,16}. A peak appears in the conductance histogram near 0.5 $G_{0}$ for Ni. 
In both cases the peak for the clean metal that represents a single metal atom contact (at 1.6 $G_{0}$ for Pt and 1.4 
$G_{0}$ for Ni) is suppressed after admission of CO. Weak features near 0.5 $G_{0}$ also appear in the conductance 
histograms for Au and Cu, although the change is much smaller than for Pt and Ni.

For Pt, the 0.5 $G_{0}$ feature is likely due to the conductance of a single CO molecule anchored between Pt 
electrodes \cite{16,17}. For Ni, Cu, and Au, it is not clear whether the new feature near 0.5 $G_{0}$ is due to the 
conductance of a single CO molecule, or whether it is due to the conductance of the metal nano wire whose 
surface is covered by CO. In both cases the conductance of the contact should vary with the type of metal 
because the electronic structure of the CO-metal bond depends strongly on the metal. Therefore, it is 
interesting to observe that a conductance feature near 0.5 $G_{0}$ is found for all metals.  

The importance of the changes in the conductance histograms for Ni, Pt, Cu, and Au decreases in that 
order. This order agrees with the adsorption properties of CO molecules on metal surfaces \cite{15}. On transition 
metal (Ni, Pt) surfaces, the electron donation from the 5$\sigma$ orbital to the metal and the back donation from the 
metal to the 2$\pi^{*}$ orbital occur simultaneously, and a strong chemical bond is formed between the CO 
molecule and the metals. The present results show that the adsorption properties on macroscopic surfaces 
reflect the adsorption properties on atomic-scale metal nano structures. On the other hand, there is a 
difference in the adsorption properties between the macroscopic surface and the atomic scale nano structure. 
While a CO molecule does not adsorb chemically on macroscopic Au surfaces, CO molecules appear to 
coordinate to Au nano contacts judging from the change in the conductance histogram (Fig.~\ref{fig1}(a)). Indeed, 
novel catalytic properties have been reported for Au nano particles \cite{18}. Theoretical calculations support the 
idea that the Au nano contact is chemically active, with strong chemisorption of O$_{2}$ and CO \cite{19}. 

\begin{figure}
\begin{center}
\leavevmode\epsfxsize=50mm \epsfbox{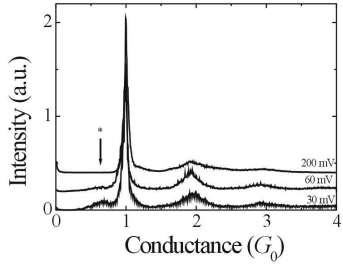}
\caption{
Bias voltage dependence of conductance histograms for Au atomic-size contacts after admitting CO 
for various values of the bias voltage.}
\label{fig2}
\end{center}
\end{figure}

\begin{figure}
\begin{center}
\leavevmode\epsfxsize=50mm \epsfbox{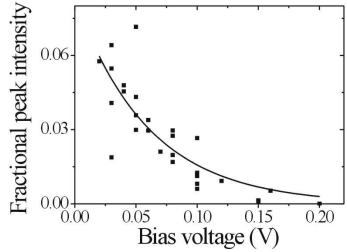}
\caption{Intensity of the fractional feature in the histograms normalized to the intensity of the peak at 1 $G_{0}$, 
plotted as a function of bias voltage for Au atomic-size contacts after admitting CO.}
\label{fig3}
\end{center}
\end{figure}

For Au nano contacts we observed systematic changes in the conductance behaviour and reproducible 
conductance traces, characteristic of Au nano contacts in the presence of CO. For this reason we concentrate 
on Au nano contacts in the following. The conductance histogram changes with bias voltage, and the change 
is continuous and reversible (Fig.~\ref{fig2}). The fractional conductance feature is clearly observed at low bias 
voltage, and its intensity decreases with bias voltage. Figure~\ref{fig3} shows the intensity of the fractional feature 
normalized to the intensity of the peak at 1 $G_{0}$, plotted as a function of the bias voltage. The fractional 
conductance feature gradually decreases and disappears at a bias voltage of about 200 mV, indicating that the 
structure having a fractional conductance is unstable at high bias voltage, suggesting that the interaction 
between the CO molecule and the Au nano contact is weak. It is noteworthy that the fractional conductance 
feature appears at low bias voltage, even after it first disappeared at high bias voltage. This indicates that the 
CO molecule is not blown off the surface but remains bound to one of the two electrodes at high bias 
voltage. 

\begin{figure}
\begin{center}
\leavevmode\epsfxsize=50mm \epsfbox{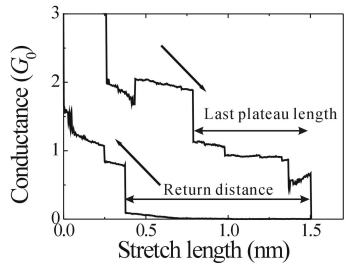}
\caption{Tycal breaking and return traces for Au after admitting CO recorded at a bias voltage of 100 mV. 
The length of the last plateau is defined as the distance between the points at which the conductance drops 
below 1.2 $G_{0}$ and 0.2 $G_{0}$, respectively.}
\label{fig4}
\end{center}
\end{figure}

In order to find evidence as to how the CO molecule is adsorbed on, or incorporated into the Au nano 
wire, we have analysed the conductance traces. Figure~\ref{fig4} shows typical breaking and return traces for Au after 
admitting CO. The stretching length was calibrated with the length histogram of the last conductance plateau 
for clean Au \cite{20}. There are three characteristics in the conductance traces. The last plateau, by which we 
mean the part having a conductance near 1 $G_{0}$, extends to quite long length; the conductance drops abruptly 
just before breaking to about 0.5 $G_{0}$; and the conductance smoothly increases with further stretching on this 
fractional conductance plateau. These characteristics are observed reproducibly and they are typical for Au 
nano contact in the presence of CO. In the following these characters are discussed.

\begin{figure}
\begin{center}
\leavevmode\epsfxsize=50mm \epsfbox{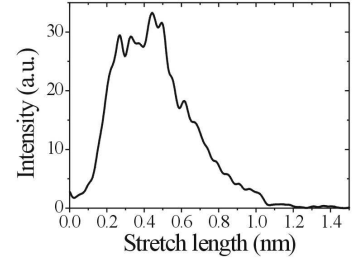}
\caption{The length histogram of the last conductance plateau as defined in Fig.~\ref{fig4}. The histogram is obtained 
from 3000 experiments recorded at a bias voltage of 100 mV.}
\label{fig5}
\end{center}
\end{figure}

\begin{figure}
\begin{center}
\leavevmode\epsfxsize=50mm \epsfbox{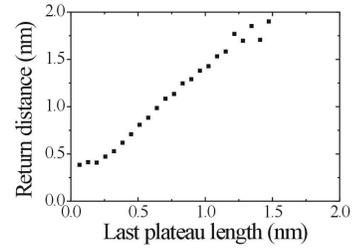}
\caption{The average of the return distance as a function of the length of the last plateau as defined in Fig.~\ref{fig4} 
recorded at a bias voltage of 100 mV. }
\label{fig6}
\end{center}
\end{figure}

First, we discuss the length of the Au nano contact. Figure~\ref{fig5} shows the length histogram of the last 
conductance plateau for Au after admitting CO. The length of the last plateau is taken here as the distance 
between the points at which the conductance drops below 1.2 $G_{0}$ and 0.2 $G_{0}$, respectively. A conductance at 1 
$G_{0}$ corresponds to a clean Au monatomic contact or chain of Au atoms. The 1$G_{0}$-plateau was observed to 
stretch up to 1 nm in length. Figure~\ref{fig6} shows the average of the return distance as a function of the length of 
the last plateau. Apart from an offset of 0.5nm due to the elastic response of the banks \cite{2}, the relation is 
approximately 1:1, suggesting that a fragile structure is formed with a length corresponding to that of the last 
plateau, which is unable to support itself when it breaks and collapses onto the banks on either side. The long 
plateaux of over 1 nm length and the 1:1 relation were also observed for clean Au monatomic wires \cite{2}, 
supporting the formation of the monatomic wire in the present case. While clean Au forms a monatomic wire 
under UHV at low temperature the results show that Au monatomic wires also form in the presence of CO.

Second, we discuss the change in conductance during the breaking process. The conductance drops 
abruptly down to about 0.5 $G_{0}$ just before breaking the contact, which produces a fractional feature in the 
conductance histogram. This conductance drop should be caused by the adsorption of CO on the Au nano 
wire or incorporation of CO into the Au wire because it is not observed for clean Au. It should be noticed 
that the wire breaks once the conductance has dropped to about 0.5 $G_{0}$. For Pt in the presence of H$_{2}$ and Ag 
in the presence of O$_{2}$, the conductance decreases stepwise after the conductance drops from 1 $G_{0}$, and the 
contact can be stretched further \cite{21,22}. This indicates that CO adsorption or incorporation destabilized the 
Au nano wire. 

The destabilization mechanism of the Au nano wire by CO we will now discuss in terms of the charge 
transfer between the CO molecule and Au, and in relation to the stabilization mechanism of the Au 
monatomic wire. The Au d band is located below the Fermi level, and thus, the electron back donation from 
the Au d band to 2$\pi^{*}$ orbital of CO does not play an important role. The donation from the 5$\sigma$ orbital of CO 
to the metal is dominant, and the electron would be transferred from CO to Au. The charge transfer between 
CO and Au was studied for Au nano particles deposited on TiO$_{2}$, which showed novel catalytic properties 
\cite{23}. Adsorption of CO on Au particles induces charge transfer from Au to TiO$_{2}$, suggesting that the electron 
is transferred from CO to Au. A monatomic wire is formed only for 5d metal as a result of the following 
mechanism. For 5 d metals, the s occupation increases at the expense of the d electrons due to relativistic 
effects in electronic structure. Since the top of the d band consists of states with anti-bonding character that 
are now partially depleted, the d bond becomes stronger. While the d electrons tend to compress the lattice, 
the s electrons exert an opposing Fermi pressure. At the surface, the spill out of the s electron cloud into the 
vacuum relieves some of the s electron pressure, and allows a strengthening of the bonds at the surface. The 
monatomic wire geometry allows for an even larger s pressure relief. Therefore, the bond in the wire is much 
stronger than the bond in the bulk, and monatomic wires are formed for 5d metals, Ir, Pt and Au \cite{3,24}. 
When CO adsorbs on the surface of the Au monatomic wire, spill out of the s electron cloud into the vacuum 
will be inhibited. Furthermore, the density of states (DOS) near the Fermi level with anti-bonding character 
will increase due to the charge transfer from CO to Au. These mechanisms will decrease the bond strength of 
the Au in the wire, and the Au monatomic wire will be destabilized by CO adsorption or incorporation.

Third, we discuss the increase in conductance with the stretch length after the conductance drops down 
to near 0.5 $G_{0}$. A similar increase in the conductance with elongation is observed for Al nano contacts \cite{25}, 
and for Au nano contacts in the presence of H$_{2}$ \cite{26}. The former is explained by the change in DOS at Fermi 
level. The elongation of the Al contact induces a narrowing the s$_{pz}$ band, leading to an increase in the DOS at 
the Fermi level, and a concomitant increase in the conductance. The latter (the effect for Au+H$_{2}$) was 
explained as due to a gradual change in the molecular orientation. Barnett et al. used density functional 
theory (DFT) to calculate the conductance of the various atomic configurations, including a system having a 
H$_{2}$ molecule incorporated into a Au monatomic wire. The conductance of the perpendicular configuration, 
for which the molecular axis is perpendicular to the contact axis, was calculated to be about 0.1 $G_{0}$, and the 
conductance of the parallel configuration increased up to 0.25 $G_{0}$ \cite{27}. Based on these calculation results the 
increase in conductance was explained by the change in configuration of hydrogen molecule. The H$_{2}$ 
molecule incorporates into the Au wire in the perpendicular configuration and the molecule turns toward an 
orientation parallel to the contact axis, which shows up as an increasing conductance as the electrodes are 
pulled apart. The present results may be explained by both a change in the DOS or a changing orientation of 
the CO molecule. Recent DFT calculation results for Pt nano contacts in the presence of CO show a slight 
increase in the conductance with the elongation \cite{17}, and this mechanism may also apply to Au in the 
presence of CO.

\section{CONCLUSIONS}
\label{sec4}

We have studied the conductance of Ni, Cu, Pt, and Au nano contacts in the presence of CO. In all cases 
a fractional conductance feature appears in the conductance histogram near 0.5 $G_{0}$ after admitting CO. The 
change in the conductance histogram for Ni, Pt, Cu, and Au decreases in that order, in agreement with the 
adsorption properties on macroscopic metal surfaces. For Au, the intensity of the fractional conductance 
feature continuously and reversibly changes with the bias voltage. The feature disappears at a bias voltage of 
200 mV, indicating that the CO-Au bond is weak. The results give evidence that Au atomic chain formation 
persists in the presence of CO, but bonding of a CO molecule weakens the chain and gives rise to a drop of 
the conductance to about 0.5 $G_{0}$.

\acknowledgments{
This work is part of the research program of the Stichting voor Fundamenteel Onderzoek der Materie (FOM), 
which is financially supported by NOW..MK has been supported by the Yamada Science Foundation.}

\end{multicols}

\begin{references}
\bibitem{20}.	The conductance traces are obtained as a function of piezo voltage. It is reported in \cite{2} that the length 
histogram of the last conductance plateau for clean Au, which is defined as having a conductance 
between 1.3 $G_{0}$ and 0.7 $G_{0}$, shows a number of equidistant maxima, at multiples of 0.257 nm \cite{3}. We use 
this to determine the ratio between the relative change in piezo voltage with respect to the stretch length, 
assuming that the inter-peak distance is 0.26 nm in the length histogram for clean Au. The stretch length 
was calibrated in this way with clean Au before each measurement of the Au in the presence of CO.
\end{references}
\end{document}